\documentclass[mathleft
]{an-fermipreprint}
\usepackage{graphicx}
\usepackage{times}
\usepackage{deluxetable}
\usepackage{threeparttable}
\newcommand{\angstrom}{\textup{\AA}}
\begin{document}


\title{Discovery of a New Blue Quasar: SDSS J022218.03-062511.1}

\author{Mees B. Fix\inst{1}, J. Allyn Smith\inst{1}\fnmsep\thanks{Corresponding author:
  \email{smithj@apsu.edu}\newline}, Douglas L. Tucker\inst{2}, William Wester\inst{2}, James
  Annis\inst{2}
}
\titlerunning{Discovery of SDSS J022218.03-062511.1}
\authorrunning{Mees B. Fix et. al}
\institute{
Department of Physics \& Astronomy, Austin Peay State University, Clarksville, TN 37044
\and 
Fermi National Accelerator Lab, Batavia, IL 60510-0500, USA
}

\received{2015 Jan 06}
\accepted{2015 Apr 21}
\publonline{2015 Aug 01 {\em(expected)}}


\abstract{%
We report the discovery of a bright blue quasar: SDSS J022218.03-062511.1.  This object was discovered
spectroscopically while searching for hot white dwarfs that may be used as calibration sources for
large sky surveys such as the Dark Energy Survey or the Large Synoptic Survey Telescope project.  We
present the calibrated spectrum, spectral line shifts and report a redshift of $z = 0.521\pm0.0015$ and 
a rest-frame $g$-band luminosity of $8.71 \times 10^{11}$ $L_{\odot}$.
} 

\maketitle

\section{Introduction}
Discovered in the 1960s, Quasi-Stellar Objects (QSOs or quasars) are among the most luminous and most
distant objects in the observable universe.  Due to these properties, quasars have been effective probes
of many facets of the high-energy and of the high-redshift Universe, including supermassive black holes
(Ghisellini et~al. 2014), the intergalactic medium (Noterdaeme et~al. 2012), the large scale structure
of the Universe (Springel et~al. 2005; Croom et~al. 2009), and the era of reionization (Becker et~al.
2001). The Sloan Digital Sky Survey (SDSS) (York et~al. 2000) and the Two-Degree Fields Survey (2dF)
(Boyle et~al. 2000) discovered thousands of these objects and enabled extensive studies of quasars
(e.g., Richards et~al. 2004; Richards et~al. 2005) and the discovery of unusual QSOs (e.g., Hall et~al.
2002).  Spectroscopic targeting of the SDSS quasars (Richards et~al. 2002 for SDSS-I/II; Ross et~al. 2012 
for SDSS-III) ensured a uniformly selected
sample of objects for study in the SDSS. That said, due to (among other things) variations in the
surface density of quasars on the sky, conflicts for fibers with other SDSS programs (e.g., galaxies and
stars), and the need to avoid fiber collisions, not all potential quasar targets were in fact assigned a
fiber by the SDSS spectroscopic targeting algorithms.

Here, we report the serendipitous discovery of the bright, blue quasar, SDSS J022218.03-062511.1.
It was missed in the original SDSS program since it was south of the footprint.  It was imaged, but 
not targeted for spectroscopy by SDSS-III spectroscopic targeting.  However, based on its SDSS colors, 
it was selected as a candidate white dwarf for our program to construct a ``Golden Sample'' of 
well-characterized, pure-hydrogen-atmosphere (``DA'') white dwarfs.  This sample is being developed to 
aid in the photometric calibration of ongoing and future Southern-sky CCD imaging survey projects such 
as the Dark Energy Survey (DES) (S\'anchez 2010; Diehl et~al. 2014) and the Large Synoptic Survey 
Telescope (LSST) (Tyson 2008).  In the following sections, we describe our targeting, observations, 
and data reductions of this ``candidate white dwarf'' (Section 2), as well as describe the the properties 
of this newly discovered quasar (Section 3).

\section{Target Selection, Observations \& Reductions}

To support the photometric calibration of the DES, we have embarked on a project to obtain a ``Golden
Sample'' of at least 30--100 well-characterized DA white dwarfs scattered
over the DES footprint to use as spectrophotometric standards.  This program is connected to the 
DES project but uses external (non-DES) observing resources to obtain imaging and spectroscopic data for
candidate DA white dwarf targets in the DES footprint. For LSST, we anticipate this sample will need to
be increased to 150-250 stars, based solely on scale-up of the survey areal coverage plans. 

The program to characterize the DAs for survey calibration draws both candidate and known white
dwarfs from several sources, notably the SuperCOSMOS white dwarf survey (Rowell \& Hambly 2011) and the
SDSS DR4 and DR7 white dwarf catalogs (Eisenstein et~al. 2006; Kleinman et~al. 2013 respectively). 
Additional targets were color selected from the SDSS DR10 (Ahn et~al. 2014) with special attention being
paid to those candidates that were in or near the planned DES SNe fields.  One of the SDSS DR10 color
selected objects was SDSS J022218.03-062511.1, which is in the DES X-2 supernova field.  This object was
noted to be fairly blue with colors corresponding to a hot white dwarf.  A search of the literature shows
this object was first identified as a blue object (PHL 1269) by Chavira (1990) and is identified on
their Figure~1.  It was later re-identified in the SDSS survey as a blue object but was not selected as 
a spectroscopic target.

We observed this target as part of our candidate white dwarf follow-up program on 2013 January 3 using
three 400 second exposures with the Dual-Imaging Spectrograph (DIS) on the 3.5-m ARC telescope at Apache
Point Observatory (APO), New Mexico.  The DIS was set to use the standard low-blue/low-red configuration
and a 2.0$\arcsec$ slit width.  For the DIS red channel we used the R300 grating, which covers 4620
$\angstrom$ at 2.31 $\angstrom$/pix resolution, and for the the blue channel we used the B400 grating,
which  covers 3660 $\angstrom$ at 1.83 $\angstrom$/pix resolution.  The combined coverage with these
two  gratings runs from $<$3600 to $>$9000 $\angstrom$. All of the data were processed using the
standard IRAF~\footnote{IRAF is distributed by the National Optical Astronomy Observatory, which is
operated by the Association of Universities for Research in Astronomy (AURA) under cooperative agreement
with the National Science Foundation.} spectroscopic packages supplemented with the  {\tt DIStools} IRAF
external package developed by Gordon Richards, which is specifically used to reduce data taken using the
DIS instrument at APO.  

\section{Results and Discussion}

\begin{figure}[h]
\includegraphics[scale=0.42]{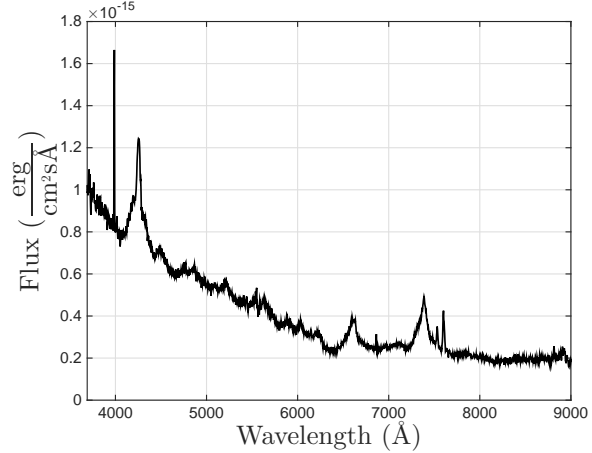}
\caption{A flux and wavelength calibrated spectrum of SDSS J022218.03-062511.1 in the redshifted frame
with labeled emission lines.  The line near 4000 $\angstrom$ is a processing artifact. \label{spectrum}}
\end{figure}

Comparing the SDSS J022218.03-062511.1 spectrum, Figure~\ref{spectrum}, to QSO spectra observed by
SDSS we determined which spectral lines were visible.  The four visible lines were identified and these
were used to calculate the associated redshift.  These values are given in Table~\ref{redshift}.  The
average redshift for this QSO is determined to be $z \approx 0.521\pm 0.0015$.  This redshift was 
verified by visual inspection and cross-comparison of the SDSS spectra of QSOs of similar redshift 
within the SDSS DR7 quasar catalog (Schneider et~al. 2010).  

\begin{table}
\centering
\caption{Calculated redshift for SDSS J022218.03-062511.1 
\label{redshift}}
\begin{tabular}{cccc}
\hline
Spectral Line & $\lambda_{\mbox{rest}}$ [$\angstrom$] & $\lambda_{\mbox{obs}}$ [$\angstrom$] & {$z$}\\
\hline
MgII & 2796 & 4256.577 & 0.522\\
H$\gamma$ & 4341.69 & 6617.334 & 0.524\\
H$\beta$ & 4862.69 & 7387.625 & 0.519 \\
OIII & 5008.24 & 7601.456  & 0.518 \\
\hline
\end{tabular}
\end{table}

\begin{figure}[h]
\includegraphics[scale=0.42]{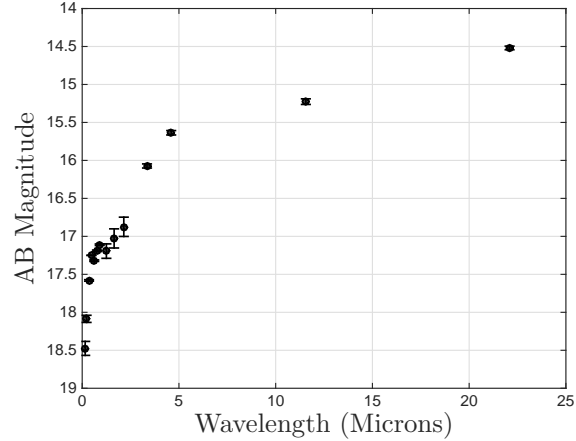}
\caption{The SED for SDSS J022218.03-062511.1 based on available data from the GALEX, SDSS, 2MASS, and
WISE archives (Table 3). AB magnitudes are used in the plot.  GALEX and SDSS use this system and the
2MASS and WISE values were converted to AB values as referenced in the Table~3 discussion. \label{QSO-mags}}
\end{figure}

A search of the NED\footnote{https://ned.ipac.caltech.edu/} database showed this object was a 
previously identified X-ray source and further, it was observed by 
ROSAT\footnote{http://www.xray.mpe.mpg.de/rosat/survey/rass-bsc/} (Voges et~al. 1999), 
GALEX\footnote{http://galex.stsci.edu/GR6/} (Martin et~al. 2005), 
2MASS\footnote{http://www.ipac.caltech.edu/2mass/releases/allsky/} (Skrutskie et~al. 2006), 
and WISE\footnote{http://wise2.ipac.caltech.edu/docs/release/allsky/} (Wright et~al. 2010). 
However until now, no spectrum was ever obtained so it remained unidentified as a quasar.  The 
USNO parallax observations (Monet et~al. 2003) are consistent with the zero proper motion expected 
of a quasar.  Table~\ref{info} lists the pertinent coordinate and ID information.
Table~\ref{mags} lists the available ROSAT, GALEX, SDSS, 2MASS, and WISE photometry.  The 2MASS 
and WISE values are also converted to AB values following Odenwald et al.~(2003) and the prescription
given on the WISE IPAC website as noted in Table~3; the errors were not changed.  As the data show, 
this is a fairly bright quasar that had avoided identification to date. Figure~\ref{QSO-mags} shows the 
far-ultraviolet to mid-infrared spectral energy distribution (SED) for this quasar based on archival 
data from GALEX, SDSS, 2MASS, and WISE.


\begin{table}
\centering
\caption{Identification information for SDSS J022218.03-062511.1 
\label{info}}
\begin{tabular}{cc}
\hline
ID/Coordinate & Value\\  
\hline
SDSS ID & J022218.03-062511.1 \\
SDSS photo ObjID & 1237679439351644175 \\
RA (J2000) & 35.57513 deg. \\
Dec (J2000) & -6.41976 deg. \\
l & 173.24232 \\
b & -60.08776 \\
\hline
RASS-6dFGS & 6dF J0222181-062511 \\
ROSAT & 1RXS J022218.0-062513 \\
GALEXASC & J022218.05-062510.2 \\
2MASS & J02221804-0625111 \\
WISE & J022218.03-062511.0 \\
\hline
\end{tabular}
\end{table}

\begin{table}
\centering
\caption{Photometric data for SDSS J022218.03-062511.1
\label{mags}}
\begin{tabular}{cccc}
\hline
Filter & value & uncertainty & unit\\
\hline

ROSAT-1RXS & 2.04 $\times 10^{-12}$ & 2.348 $\times 10^{-13}$ & erg/cm$^{2}$/s \\
GALEX-FUV & 18.4747 & 0.0920 & AB-mag \\
GALEX-FUV & 18.7425 & 0.1099 & Vega-mag \\
GALEX-NUV & 18.0863 & 0.0465 & AB-mag \\
GALEX-NUV & 17.6210 & 0.0347 & Vega-mag \\
SDSS-u & 17.58 & 0.01 & AB-mag \\
SDSS-g & 17.25 & 0.00 & AB-mag \\
SDSS-r & 17.32 & 0.01 & AB-mag \\
SDSS-i & 17.19 & 0.01 & AB-mag \\
SDSS-z & 17.11 & 0.01 & AB-mag \\
2MASS-J & 17.194 & 0.095 & AB-mag\textsuperscript{*} \\
2MASS-J & 16.354 & 0.095 & Vega-mag \\
2MASS-H & 17.027 & 0.126 & AB-mag\textsuperscript{*} \\
2MASS-H & 15.657 & 0.126 & Vega-mag \\
2MASS-K & 16.874 & 0.127 & AB-mag\textsuperscript{*} \\
2MASS-K & 15.034 & 0.127 & Vega-mag \\
WISE-W1 & 16.073 & 0.027 & AB-mag\textsuperscript{*} \\
WISE-W1 & 13.39 & 0.027 & Vega-mag \\
WISE-W2 & 15.636 & 0.029 & AB-mag\textsuperscript{*} \\
WISE-W2 & 12.317 & 0.029 & Vega-mag \\
WISE-W3 & 15.226 & 0.037 & AB-mag\textsuperscript{*} \\
WISE-W3 & 9.984 & 0.037 & Vega-mag \\
WISE-W4 & 14.519 & 0.024 & AB-mag\textsuperscript{*} \\
WISE-W4 & 7.915 & 0.024 & Vega-mag \\

\hline

\end{tabular}
	\begin{tablenotes}
        \item{*}\footnotesize{The 2MASS and WISE values were converted to AB values following 
        Odenwald et al.~(2003) and the prescription given on the WISE IPAC website 
        (wise2.ipac.caltech.edu/docs/release/prelim/expsup/sec4\_3g.html)}
    \end{tablenotes}

\end{table}

\begin{figure}[h]
\includegraphics[scale=0.45]{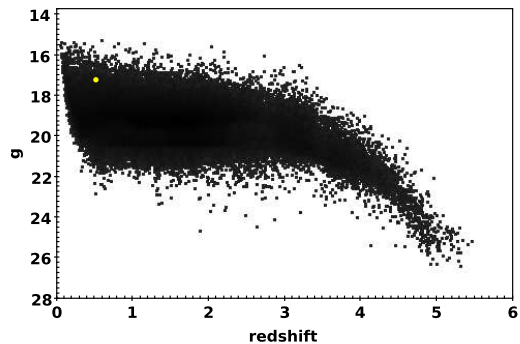}
\caption{The SDSS DR7 quasar catalog redshifts plotted versus their $g$-magnitude. SDSS
J022218.03-062511.1 is shown as the dot to show its relation to the other 100,000 objects in the
catalog. \label{redshift-g}}
\end{figure}

\begin{figure}[h]
\includegraphics[scale=0.45]{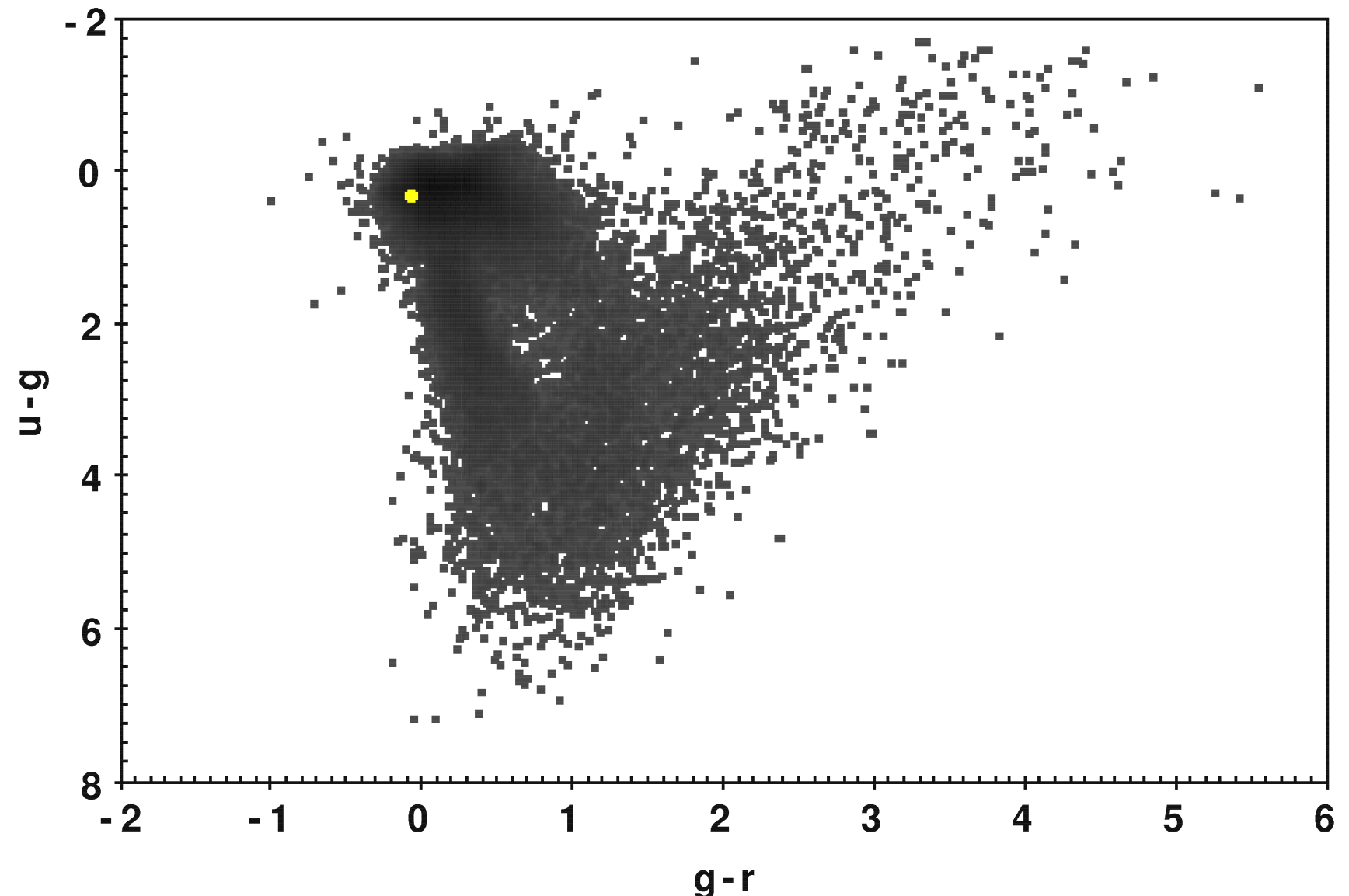}
\caption{The SDSS DR7 quasars shown in $(u-g)$ vs. $(g-r)$ space. SDSS J022218.03-062511.1 is 
shown as the dot to show its relation to the catalog data. \label{ug-gr}}
\end{figure}

\begin{figure}[h]
\includegraphics[scale=0.45]{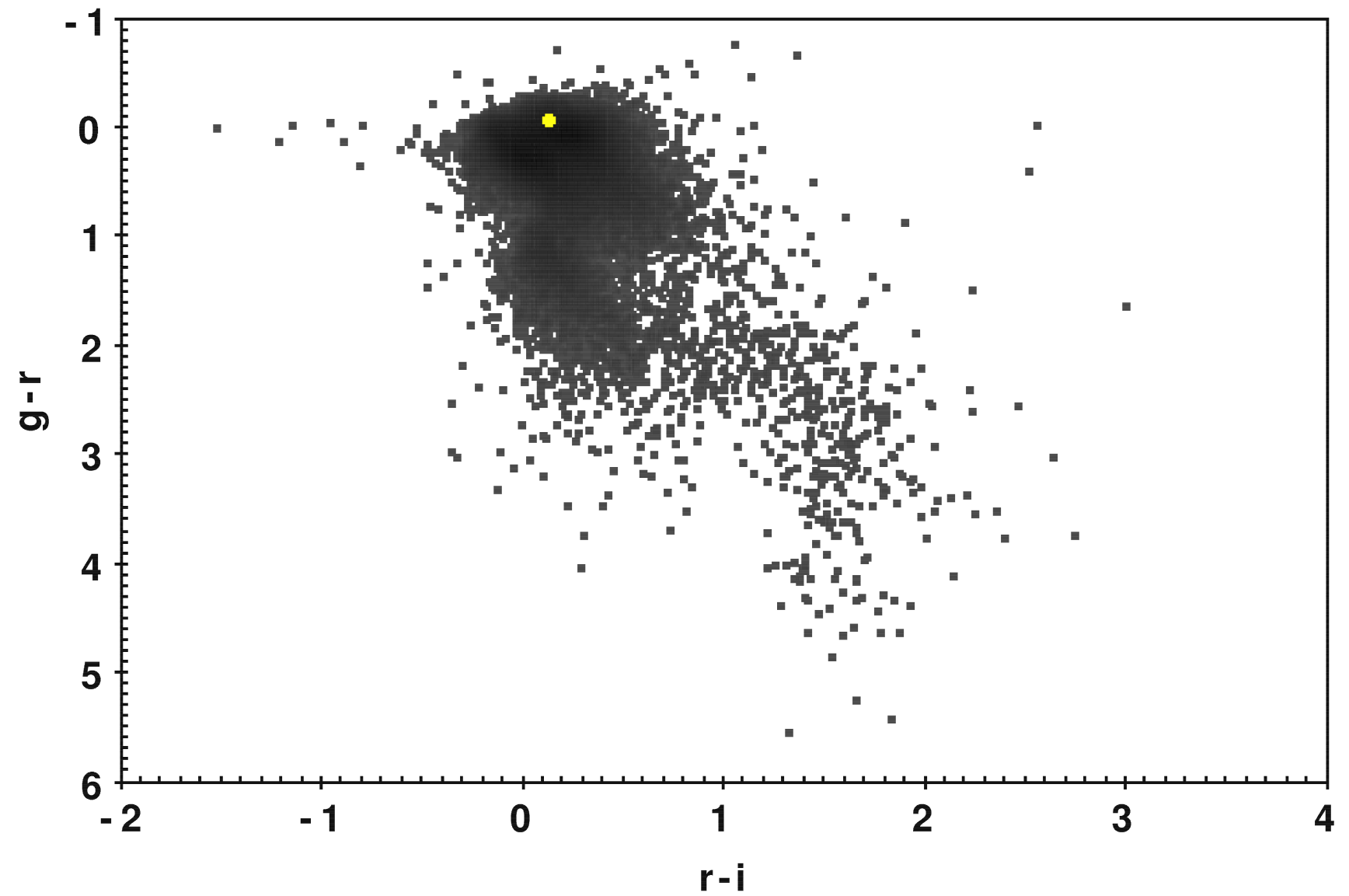}
\caption{The SDSS DR7 quasars shown in $(g-r)$ vs. $(r-i)$ space. SDSS J022218.03-062511.1 is 
shown as the dot to show its relation to the catalog data. \label{gr-ri}}
\end{figure}

A comparison to the SDSS DR7 quasar catalog (Schneider et~al. 2010) shows this object would fall in the
brightest 1.1\% of the SDSS quasars based on its $g$-band (apparent) magnitude.  It is significantly
brighter than the average SDSS DR7 quasar in the same redshift range, based on an average of 213 objects
in our redshift plus or minus the uncertainty.  While this is not the brightest of these quasars, it
would be one of the brightest 3--6 in each SDSS filter in the redshift range had it been included in the
DR7 catalog.  To illustrate the comparison to the SDSS DR7 catalog we present Figure~\ref{redshift-g} 
to show how this compares with the 100,000$+$ DR7 quasars.  We also show in Figure~\ref{ug-gr} and 
Figure~\ref{gr-ri} where this quasar falls in observed color-color space. Other than being bright, though, 
this quasar does not appear to be remarkable in any other manner. 

A search of other quasar catalogs derived from the SDSS data show this object would still avoid 
detection given the search parameters used in developing those catalogs.  Abraham et~al. (2012) 
searched the SDSS-DR7 data set, so this object had not yet been observed.  Bovy et~al. (2011) searched
DR8 but used a bright cutoff of $i = 17.75$: this object would be too bright to be included in their
study.  Ross et~al. (2012) searched the area where this quasars is found, but it was too close in
redshift to be included.  

Finally, we derive the $g$-band luminosity of SDSS J022218.03-062511.1.  The luminosity distance for
this redshift $z=0.521$ quasar is $D_{L}= 3008.7$ Mpc ($H_{0}=69.6 \pm 0.7$ km/s; $\Omega_{M} = 0.286
\pm 0.008$; $\Omega_{Vac}$ = 0.714; assuming a flat Universe) (Wright 2006, Bennett et al.
2014).\footnote{http://www.astro.ucla.edu/$\sim$wright/CosmoCalc.html}  The $g$-band k-correction (Oke
\& Sandage 1968), calculated via synthetic photometry of the rest-frame and the observed-frame ARC 3.5m
spectrum, is $k_{g}= -0.511$ mag.  (We are able to calculate the k-correction for the $g$-band directly
from the ARC-3.5m spectrum because we have complete spectral coverage for $g$-band in both the observed
and the transformed rest frames.  The $r$- and $i$-bands however, do not provide complete coverage in
both the observed and transformed rest frames.)  The $g$-band interstellar extinction, taken from the
SDSS DR12 CAS\footnote{http://skyserver.sdss.org/dr12}, is $A_{g} = 0.101$ mag.  Taking these values, 
we compute an interstellar extinction corrected $g$-band absolute magnitude for this quasar of 
$M_{g} = -24.73$, which is equivalent to a rest-frame g-band luminosity of 
$8.71 \times 10^{11}$ $L_{\odot}$ (where the $g$-band absolute magnitude of the Sun is $+5.12$).

\acknowledgements{
Partial support for MBF and JAS was provided by the Department of Energy Visiting Faculty Program
run by the Department of Energy Office of Science.  Additional support came from the Fermilab
Center for Particle Astrophysics.  Based on observations obtained with the Apache Point Observatory
3.5-meter telescope, which is owned and operated by the Astrophysical Research Consortium. {\tt
DIStools} is used for spectral reductions at Apache Point and was developed by Gordon Richards
while at the University of Chicago. 

Funding for SDSS-III has been provided by the Alfred P. Sloan Foundation, the Participating
Institutions, the National Science Foundation, and the U.S. Department of Energy Office of Science.
The SDSS-III web site is http://www.sdss3.org/.

SDSS-III is managed by the Astrophysical Research Consortium for the Participating Institutions of
the SDSS-III Collaboration including the University of Arizona, the Brazilian Participation Group,
Brookhaven National Laboratory, Carnegie Mellon University, University of Florida, the French
Participation Group, the German Participation Group, Harvard University, the Instituto de
Astrofisica de Canarias, the Michigan State/Notre Dame/JINA Participation Group, Johns Hopkins
University, Lawrence Berkeley National Laboratory, Max Planck Institute for Astrophysics, Max
Planck Institute for Extraterrestrial Physics, New Mexico State University, New York University,
Ohio State University, Pennsylvania State University, University of Portsmouth, Princeton
University, the Spanish Participation Group, University of Tokyo, University of Utah, Vanderbilt
University, University of Virginia, University of Washington, and Yale University. 

This publication makes use of data products from the Two Micron All Sky Survey, which is a joint
project of the University of Massachusetts and the Infrared Processing and Analysis
Center/California Institute of Technology, funded by the National Aeronautics and Space
Administration and the National Science Foundation.

This research has made use of the NASA/IPAC Extragalactic Database (NED), which is operated by the
Jet Propulsion Laboratory, California Institute of Technology, under contract with the National
Aeronautics and Space Administration. This research has made use of the SIMBAD database, operated
at CDS, Strasbourg, France. This research made use of data from the GALEX mission.  GALEX is a NASA
small explorer, launched in 2003 April. It is operated for NASA by Caltech under NASA contract
NAS5-98034.  This publication makes use of data products from the Wide-field Infrared Survey
Explorer, which is a joint project of the University of California, Los Angeles, and the Jet
Propulsion Laboratory California Institute of Technology, funded by the National Aeronautics and
Space Administration.

The TOPCAT software package\footnote{http://www.starlink.ac.uk/topcat/} was used in much of the
plotting and analysis in this work.
}

\newpage

\end{document}